\documentclass[journal]{IEEEtran}
\usepackage{hyperref}
\usepackage{mdframed}
\usepackage{lipsum}
\usepackage{amsmath}
\usepackage{amsfonts}
\usepackage[font=small]{caption}
\usepackage{ifpdf}
\usepackage{todonotes}
\usepackage{booktabs} 
\usepackage{url}
\usepackage{pifont}
\usepackage{color}
\usepackage{tcolorbox}
\usepackage{balance}
\usepackage{makecell}
\usepackage{verbatim}

\usepackage{algorithmic}
\usepackage{array}

\usepackage{bussproofs}
\usepackage{lipsum} 
\usepackage{mathpartir}

\usepackage{caption}
\usepackage{multirow}
\usepackage{float}
\usepackage{mwe}
\newcommand{\code}[1]{\texttt{#1}}

\usepackage{url}
\usepackage{todonotes}
\newcommand{\cmark}{\ding{51}}%
\newcommand{\xmark}{\ding{55}}%

\usepackage{subfigure}
\usepackage{tikz}
\usepackage{xcolor}

\usepackage{booktabs,caption,fixltx2e}
\usepackage[flushleft]{threeparttable}

\hypersetup{
 colorlinks=true,
 linkcolor=blue,
 citecolor=blue,
 filecolor=black,
 urlcolor=black,
 pdfauthor = {},
 pdftitle = {},
 pdfkeywords = {}
}

\usepackage{listings}
\usepackage{graphicx} 
\usepackage{algorithm2e}
\usepackage{xcolor}

\SetAlFnt{\footnotesize}

\SetAlCapSty{xAlCapSty}

\SetCommentSty{xCommentSty}

\SetNlSty{mynlfont}{}{} 

\LinesNumbered

\SetSideCommentRight

\DontPrintSemicolon

\RestyleAlgo{algoruled}

\begin{document}


\title{A Comparative Study of Full Apps and Lite Apps for Android}

\author{
    Yutian Tang,~\IEEEmembership{Senior Member,~IEEE,}
    Xiaojiang Du,~\IEEEmembership{Fellow,~IEEE}%
    \thanks{Yutian Tang is with the School of Computing Science, University of Glasgow, United Kingdom (email: Yutian.Tang@glasgow.ac.uk).}%
    \thanks{Xiaojiang Du is with the Department of Electrical and Computer Engineering, Stevens Institute of Technology, United States (email: xdu6@stevens.edu).}%
}

\maketitle

\begin{abstract}
    App developers aim to create apps that cater to the needs of different types of users. This development approach, also known as the ``one-size-fits-all'' strategy, involves combining various functionalities into one app. However, this approach has drawbacks, such as lower conversion rates, slower download speed, larger attack surfaces, and lower update rates. To address these issues, developers have created ``lite'' versions to attract new users and enhance the user experience. Despite this, there has been no study conducted to examine the relationship between lite and full apps. To address this gap, we present a comparative study of lite apps, exploring the similarities and differences between lite and full apps from various perspectives. Our findings indicate that most existing lite apps fail to fulfill their intended goals (e.g., smaller in size, faster to download, and using less data). Our study also reveals the potential security risks associated with lite apps.
\end{abstract}


\section{Introduction}


Over these years, mobile devices and applications have become a ubiquitous part of our daily lives. According to AppBrain statistics \cite{appbrain}, there are over 2.68 million Android apps available on the Google Play Store. To meet the needs of different types of users, developers are continually striving to build apps with rich features. This often leads to apps that are packed with various features and libraries to support different application binary interfaces (ABIs) for different platforms \cite{Huang:2017,Tang:2021}. In this context, a feature is defined as a specific functionality in an app that satisfies a particular requirement. For example, the ability to ``pay a bill'' in a bank app can be considered a feature as it satisfies the transaction needs of a bank app user.

Combining all features into an app, also known as the ``one-size-fits-all'' strategy, can offer convenience and suitability for different types of users and devices.  However, such a ``one-size-fits-all'' strategy can bring various potential defects and risks to the target systems, including lower conversion rates, slower download speeds, larger attack surface, and lower update rates \cite{Bhattacharya:13}. To combat the software bloat in Android apps, two types of solutions are proposed \cite{Tang:2021}: first, \textit{pruning-based debloating approach} removes unwanted features from the app to build a debloated app \cite{Tang:2021,Jiang:18}. The debloated app has fewer features than the full app. For example, developers can leverage the pruning-based delobating approach to remove an out-of-date module from an app. There are two representative works in this field: RedDroid \cite{Jiang:18} performs debloating in two aspects: (1) \underline{code redundant debloating}: RedDroid regards ``dead code'' as redundant and removes them; and (2) \underline{ABI redundant debloating}: RedDroid considers ABIs that are not suitable for the target CPU architecture as redundant and removes them from the app. For example, if the CPU architecture is ARMv7, then ABIs for ARMv5, X86, MIPS(64), ARMv8, X86, X86 64 are pruned. Different from RedDroid, XDebloat allows developers to prune unwanted features from an app \cite{Tang:2021}. Specifically, developers can first annotate features they intend to prune. Then, XDebloat prunes them to build a debloated app. Second, \textit{module-based debloating approach} allows developers to organize apps into modules. Thus, developers can download and install modules in an on-demand manner. One representative framework is the App Bundle Framework \cite{appbundle} offered by Google. With the app bundle framework, an app bundle is organized by one base module and zero or more feature modules. When an app bundle is downloaded, its base module is downloaded and its feature modules are downloaded on demand. 

\noindent\textbf{Motivation}. Nevertheless, automating the process of debloating an app presents its fair share of challenges. Developers and researchers face several challenges when debloating apps. \textbf{C1} The uniqueness of Android apps: Unlike traditional software, features in an app can vary due to the unique nature of the Android system. For example, Android apps interact with users through UI elements and have more entry points than traditional software. \textbf{C2} Defining features in apps: Compared to traditional software, more factors need to be considered when defining features in an app, such as permissions, hardware, and UI. \textbf{C3} UI-related issues: Pruning apps is different from pruning traditional software as UI-related issues must be taken into account during the pruning.

\noindent\textbf{Goal of this study.} In this study, we aim to gain insights from the practices of real-world app developers by examining how they create ``lite'' versions of their apps. Normally, a lite app is defined as an app that has a limited set of features in comparison to its full version. These lite apps are developed by official app developers and can be considered as a form of debloating in comparison to the full version of the app. Understanding the methods and considerations used by developers in creating lite apps can (1) assist researchers in developing effective debloating approaches and tools; and (2) assist developers in creating their own lite apps by refactoring their legacy apps.

\noindent\textbf{Contribution.} In summary, we make the following contributions in this paper:

\noindent$\bullet$ In this paper, we conduct a \emph{first} comparative study of lite apps and their counterpart full apps. We collect 260 lite-full app pairs from Google Play;

\noindent$\bullet$ We conduct a systematic quantitative and qualitative analysis on the app pairs from various aspects, including non-code similarity, code, similarity, permission, privacy policy, resource consistency, performance improvement, features included and excluded in lite apps, and potential security concerns in lite apps; and

\noindent$\bullet$ The results of this study shed a light on understanding the existing practices of building lite apps, and show the limitations and potential drawbacks of existing lite apps.

\section{Motivation Example}

Investigating lite apps can offer valuable insights for researchers and developers. Lite apps are designed to be a compact version of their full app counterparts, and comprehending the design choices and functional disparities between the two can assist in the development of debloating techniques and tools. This study utilizes the example of LinkedIn and its LinkedIn Lite app to illustrate the design of lite apps and the variations between them and full apps.

\subsection{Meta Comparison}

\begin{table}[!htpb]
\centering
\vspace{-1em}
\caption{LinkedIn App vs. LinkedIn Lite App}
\vspace{-1em}
\scalebox{0.9}{
\begin{tabular}{cccc}
\hline
& \textbf{LinkedIn} & \textbf{LinkedIn Lite} & \textbf{Similarity} \\ \hline
\textbf{Apk Size}    &  80.1 MB  & 8.1 MB &  / \\ \hline
\textbf{\# Permissions} &   41  &  27  &  100\% \\ \hline
\textbf{\# Activity}  &  25     &  14  &  50\%  \\ \hline
\textbf{\# Content Provider}    &    9   &  3 &  0\%  \\ \hline
\textbf{\# Service}    &   23    &    14  &  42.85\% \\ \hline
\textbf{\# Broadcast Receiver}   &   17    &  14&  50\%   \\ \hline
\textbf{\# Methods}  &   56163    &  7066 &  28.93\%     \\ \hline
\textbf{\# Resources files}  &  17031  &    187&  1.37\%      \\ \hline
\end{tabular}
}
\label{tab:linked_linkedlite}
\end{table}

LinkedIn is an online service focused on employment, offering job seekers opportunities and allowing companies to post job listings. To gain an understanding of the two types of apps, we compare the APK files of LinkedIn and its Lite app. Table \ref{tab:linked_linkedlite} presents the differences between the two apps in 8 aspects including Apk size, number of permissions, number of \code{Activity}, number of \code{Content Provider}, number of \code{Service}, number of \code{Broadcast Receiver}, number of methods, and number of resource files. Furthermore, we determine the similarity between the two apps using Eq. 1, as shown in Table \ref{tab:linked_linkedlite}. As displayed, all permissions used in the Lite app are included in the LinkedIn app, with 100\% of permissions, 50\% of \code{Activities}, 42.85\% of \code{Services}, 50\% of \code{Broadcast Receivers}, 28.93\% of methods, and 1.3\% of resource files present in both apps.

\subsection{Functionality Comparison}

\begin{figure}[!htpb]
	\centering
	\includegraphics[width=0.5\textwidth]{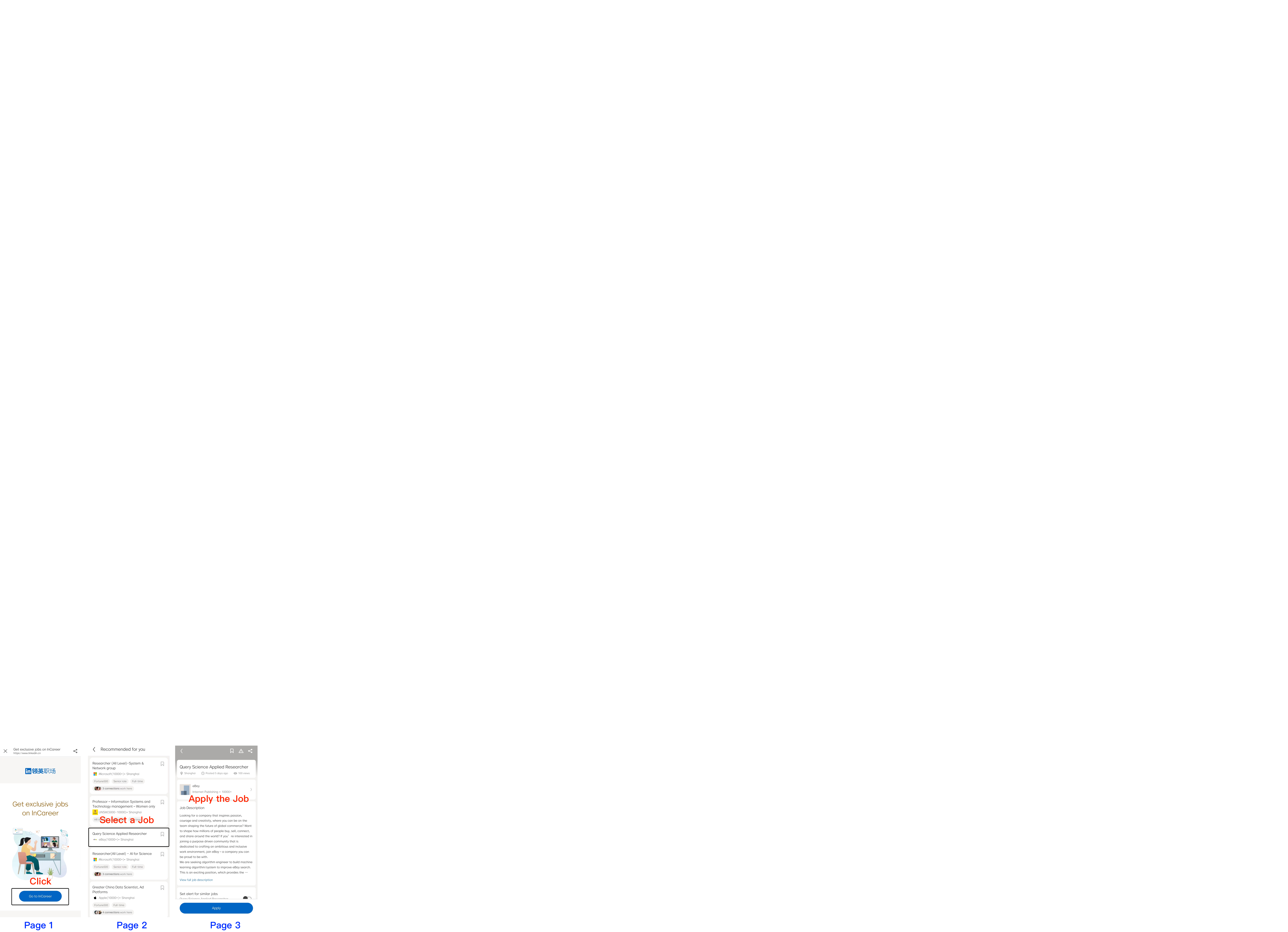}
	\caption{UI and Basic Workflow of LinkedIn Lite App}
    \label{fig:linkedinlite_workflow}
\end{figure}

\begin{figure}[!htpb]
	\centering
	\includegraphics[width=0.5\textwidth]{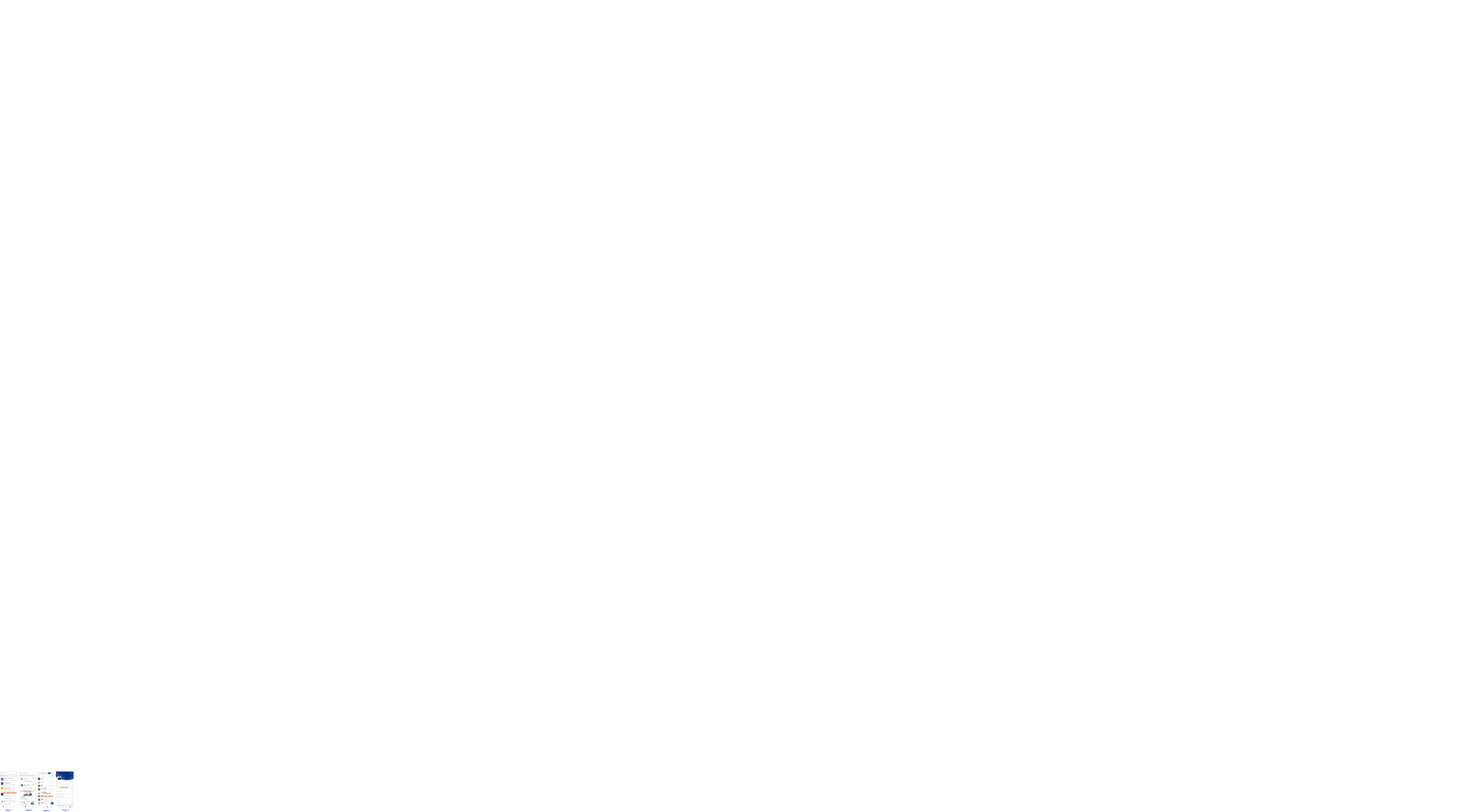}
	\caption{UI and Basic Workflow of LinkedIn App}
	\label{fig:linkedin_workflow}
\end{figure}

We perform a functionality comparison between LinkedIn app and its lite app. Fig. \ref{fig:linkedinlite_workflow} shows the main UI and basic workflow of the lite app. As shown, after logging in, users can apply for recommended jobs using pre-saved resumes. In contrast, Fig. \ref{fig:linkedin_workflow} illustrates the main user interface and functions of the LinkedIn app. It includes four main features: \textbf{job application}, \textbf{discover}, \textbf{messaging (chat and notifications)}, and \textbf{account management}. Notably, the job application feature is the same in both the lite and full versions, with the only difference being that users can select recommended jobs in the full version, while job searching is not allowed in the lite version. The \textbf{discover} feature (second tab) offers suggestions for ``people you may know'' and ``news from companies''. The \textbf{messaging} feature (third tab) allows users to send messages to their connections and receive notifications. Lastly, the \textbf{account management} feature (last tab) allows users to upload resumes, manage their account, set preferences, view saved jobs, and view applied jobs among other options.

\begin{table}[!htpb]
\centering
\caption{Functionality Comparison between LinkedIn and  Lite}
\scalebox{0.9}{
\begin{tabular}{ccc}
\hline
\textbf{Features}           & \textbf{LinkedIn} & \textbf{LinkedIn Lite} \\ \hline
\textbf{Login}  &  \cmark   & \cmark  \\ \hline
\textbf{Logout} &  \cmark  &  \xmark    \\ \hline
\textbf{Job Apply} &  \cmark &  \cmark    \\ \hline
\textbf{Job View} &   \cmark  & \cmark (Recommended job only)      \\ \hline
\textbf{Job Search} &   \cmark &  \xmark    \\ \hline
\textbf{Job Share/Save} &   \cmark & \cmark    \\ \hline
\textbf{Discover} &   \cmark & \xmark    \\ \hline
\textbf{Message(Chat,Notification)} &   \cmark & \xmark  \\ \hline
\textbf{Account Management} &   \cmark & \xmark   \\ \hline
\textbf{Preferences} &   \cmark & \xmark   \\ \hline
\textbf{Resume Management} &   \cmark & \xmark   \\ \hline
\end{tabular}
}
\label{tab:linkedIn_linkedInLite_Compare}
\end{table}

The LinkedIn and Lite apps offer different features to cater to the needs of different users. The LinkedIn offers a wide range of features such as job search, messaging, and account management, while the LinkedIn Lite only offers the core feature of job application. As shown in Table \ref{tab:linkedIn_linkedInLite_Compare}, the lite app only contains the key feature of job application, compared to the full app. Developers have intentionally limited the features in the lite app to provide a lightweight version for users who only need the core functionality. Additionally, the UI consistency between the two apps is maintained to effectively reduce the learning costs for users and allow them to apply for recommended jobs in a timely manner.

\section{Experiment and Evaluation}

\subsection{Research Questions}

In this paper, we intend to answer the following research questions (RQ). 

\noindent$\bullet$ \textbf{RQ1 (Non-code Similarity):} How similar are the disassembled permissions/resource files between lite apps and full apps?

\noindent$\bullet$ \textbf{RQ2 (Code Similarity):} To what extent is the code of lite apps similar to their full apps?

\noindent$\bullet$ \textbf{RQ3 (App Descriptions):} Whether lite apps and their corresponding full apps can have dramatically difference app descriptions?

\noindent$\bullet$ \textbf{RQ4 (Permission, Privacy Policy Consistency):} How do lite apps' claimed permissions, privacy policies stay consistent?

\noindent$\bullet$ \textbf{RQ5 (Performance Improvement with Lite Apps)}: Whether lite apps bring performance improvement comparing to full apps?

\noindent$\bullet$ \textbf{RQ6 (Shared, Included and Excluded Features in Lite Apps)}: What are the functionalities are shared, included and excluded in lite apps?

\noindent$\bullet$ \textbf{RQ7 (Potential Secure Concerns in Lite Apps)}: What are the potential secure concerns (permission, privacy) in lite apps?

\subsection{Data Collection}
To our knowledge, there is no publicly available dataset for lite and full app pairs for our study. Therefore, we created our own dataset through the following steps: 1) randomly selecting 400,000 apps from Google Play \cite{googleplay}, 2) grouping the apps by developers/companies to ensure that the lite and full app pairs are from the same developer/company, and 3) grouping the lite and full apps by comparing their app descriptions and using a similarity rate of 85\%. This resulted in 299 app pairs, with 39 of them being paid apps. After removing these paid apps, we were left with 260 app pairs.

\subsection{RQ1: Non-Code Similarity}
\noindent\textbf{Motivation.} In this RQ, we intend to evaluate the non-code similarity between lite apps and full apps. 

\noindent\textbf{Methodology.} We first investigate the similarity of lite app and full app pairs in terms of non-code assets. Specifically, we compare the following aspects: (1) declared permissions; (2) resource files: images, videos, audios, libraries, and other resource files. 

(1) Declared permissions: App permission system is a front-line mechanism to protect the privacy of Android users. If an app intend to use certain sensitive data or access to hardware resources, its have to claim certain permissions from the Android OS. If and only if such permission is granted, the app can access to certain data or hardware service. We leverage Android Asset Packaging Tool (AAPT) \cite{AAPT} to obtain the declared permissions in an app; and
 
(2) Resource files: An app can leverage different types of multimedia resources, such as images, videos, audio. In general, we compare file names and the hash value of two files with the same name in the lite version and full version.

\begin{figure}[!htpb]
	\centering
	\includegraphics[width=0.5\textwidth]{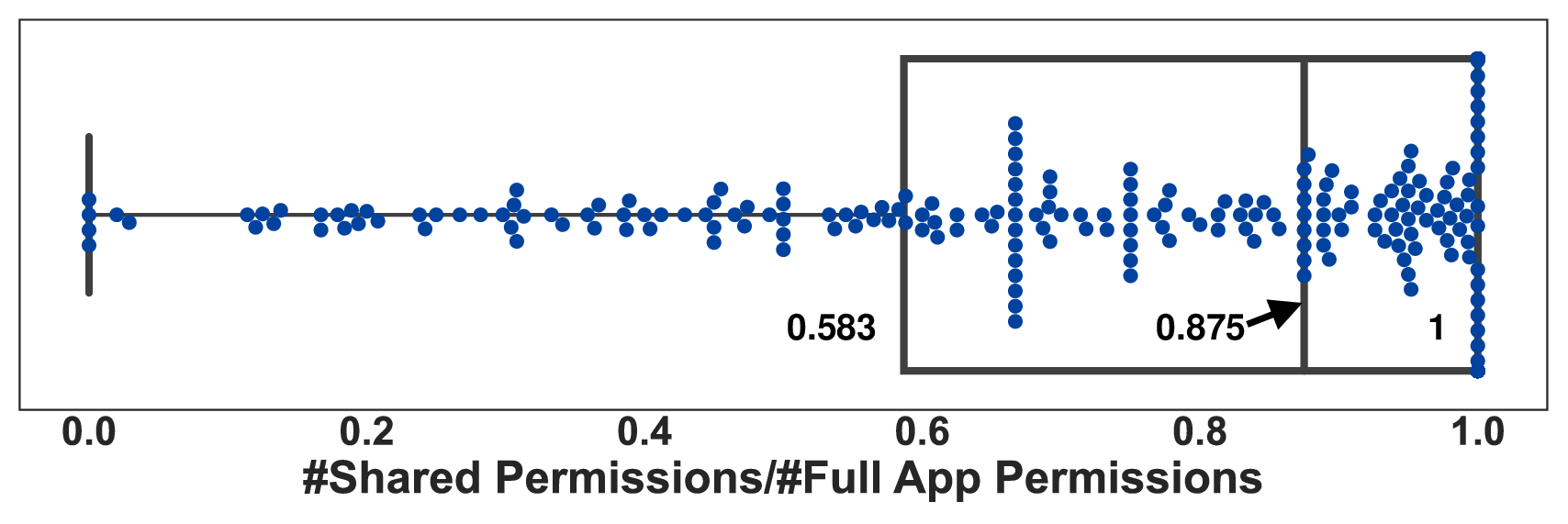}
	\caption{Boxplot of Permission Reuse Ratio}
	\label{fig:permission_reuse_rate}
\end{figure}

\noindent\textbf{Results.} As shown in Fig. \ref{fig:permission_reuse_rate}, 50\% of lite apps reuse 87.5\% permissions from their full apps; 75\% apps reuse more than 58.3\% permissions. 

The high reuse ratio of permission benefits in (1) consistency: by reusing the same permissions, the behavior and functionality of the lite apps can be consistent with the full apps, which can improve the user experience; and (2) simplicity: the development process can be simplified, as the developer does not need to consider which permissions are needed for the lite app version. However, the high reuse ratio can also bring some drawbacks, such as, privacy concerns and security concerns. We discuss this issue in RQ 3.

We also compare which kinds of permissions are removed when building lite apps. The top 10 removed permissions are: \newline
\code{gms.permission.AD\_ID} (44,16.9\%),
\code{vending.BILLING} (38,14.6\%), \newline
\code{FOREGROUND\_SERVICE} (35,13.5\%),
\code{RECEIVE\_BOOT\_COMPLETED} (34,13\%), \newline
\code{ACCESS\_COARSE\_LOCATION} (29,11.1\%), \newline
\code{SYSTEM\_ALERT\_WINDOW} (29,11.1\%), \code{CAMERA} (28,10.7\%), \newline
\code{POST\_NOTIFICATIONS} (28,10.7\%), \code{READ\_PHONE\_STATE} (27,10.3\%), and \newline\code{ACCESS\_WIFI\_STATE} (26,10\%). 


After inspecting the lite apps that these permissions are pruned, we find the potential concerns for removing these permissions are:

\noindent(1) \textbf{Performance concerns and data consumption}: The top-1 removed permission is \code{gms.permission.AD\_ID}, which is related to Google Ad. The top-2 removed permission is used for in-app purchase. By removing data consumption operations and services, developers saves computing resources in the lite apps;

\noindent(2) \textbf{Service (notification)}: Developers also prefer to set services (e.g., foreground service, geo-service) in the full app as these services need long-running operations in the background. Such long-running operations can be inappropriate to lite apps, as this may be  contrary to the original intention of lite apps; and

\noindent(3) \textbf{Network Requirement}: Developers also pay attention to the network requirements for lite app users. We find that developers use these permission to detect whether the target devices are connected to WIFI. For example, the full app \code{com.amazon.klite} checks the WIFI status to determine when to upload the crash reports to the remote server.


For the resource files \textbf{reuse} perspective, we use the \code{Similarity} metric defined in SimiDroid. In SimiDroid, the extracted key/value mapping pairs (map1,map2) are then compared based on four metrics: (1) \textbf{identical}, which indicates an exact match between compared key/value pairs; (2) \textbf{similar}, which indicates the compared pairs have the same key but different value; (3) \textbf{new}, which indicates key exists only in map1 (i.e., the component/method only exists in the lite version); and (4) \textbf{deleted}, which indicates the key only exists in map2 (i.e., the component/method only exists in the full version). The similarity score is calculated as

\begin{equation}
    similarity=max\left \{ \frac{identical}{total-new} ,\frac{identical}{total-deleted}\right \},
\end{equation}

where, total=identical+similar+new+deleted. 

\begin{figure}[!htpb]
	\centering
	\includegraphics[width=0.5\textwidth]{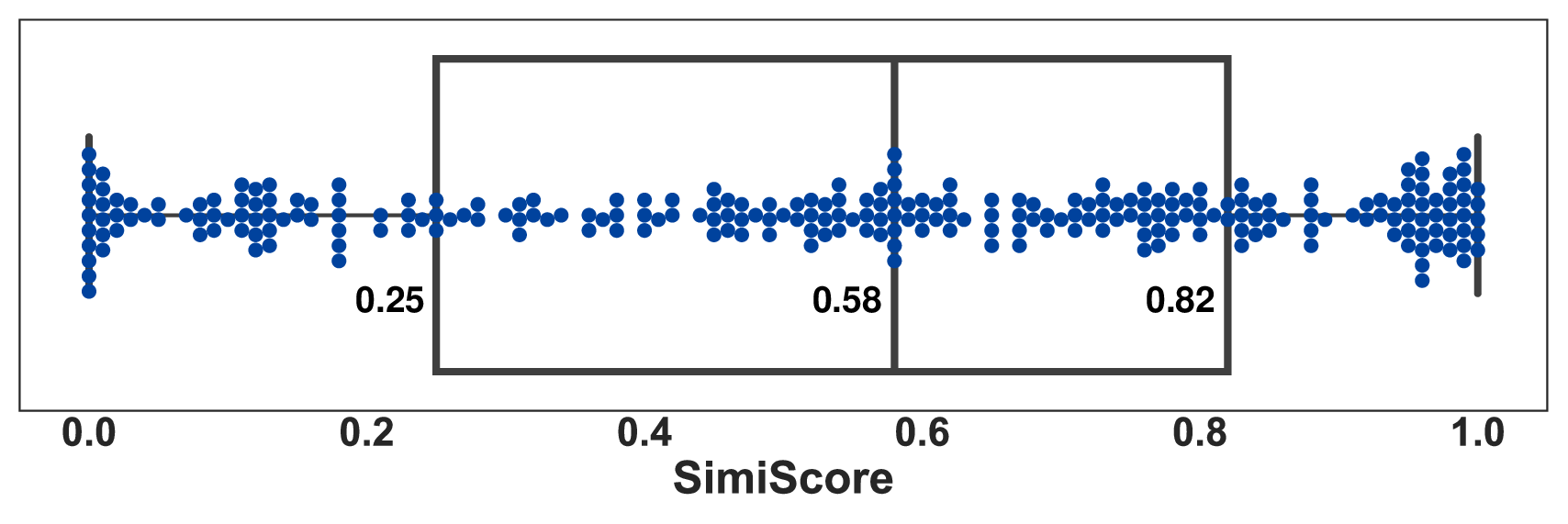}
 
	\caption{Boxplot of Resource SimiScore}
 
	\label{fig:resource_reuse_rate}
\end{figure}

%
%

As a result, the metric \code{similarity} (i.e., SimiScore) can be used to compute the similarity between two apps. On average, 54.27\% resource files in lite apps are reused from full apps. As the boxplot of SimiScore shown in Fig. \ref{fig:resource_reuse_rate}, the medium value is 0.58, and the interquartile range is 0.25 to 0.82. To understand the motivation, we manually inspect 3 app pairs for each category. We find that it is mainly due to the design choices and concerns for lite apps. Different developers may have different design choices and concerns for building lite apps. Some developers only intend to cover a few feature in the lite apps and leave most features in the full apps. While some developers intend to refine and redesign some features in the lite apps. As a result, the reuse rates for apps can vary dramatically.

\begin{tcolorbox}[boxrule=1pt,boxsep=1pt,left=2pt,right=2pt,top=2pt,bottom=2pt,title=RQ1: Non-Code Similarity]
$\bullet$ The potential concerns for removing these permissions in the lite apps are (1) performance concerns and data consumption; (2) service (notification); and (3) network Requirement;

$\bullet$ On average, the apps in the following categories have a reuse rate over 50\%: Communication, Social, Game, Map, Tools, Videos, Education, Entertainment, Life, Productivity, Personalization, Traveling, Dating;

\end{tcolorbox} 

\subsection{RQ2: Code Similarity}

\noindent\textbf{Motivation.} Typically, lite apps and full apps have similar functionalities, but due to different goals for each app version, understanding the similarities and differences in code between a lite app and a full app can provide developers and researchers with a deeper understanding of the relationship between the two.

\noindent\textbf{Methodology.} We leverage SimiDroid \cite{Li:2017}, which compares the pairwise similarities and differences between Android apps. Specifically, SimiDroid is built upon Soot, which transforms the source code into an intermediate representation named Jimple \cite{Soot}. The metric \code{similarity} can be used to compute the similarity between two apps in terms of code.


\begin{table}[!htpb]
\centering
\caption{Average Number of Components in Lite Apps and full Apps}
\begin{tabular}{c|cccc}
\hline
\multicolumn{1}{c|}{\textbf{}} & \textbf{Activity} & \textbf{Service}     & \textbf{\begin{tabular}[c]{@{}c@{}}Broadcast\\ Receiver\end{tabular}} & \textbf{\begin{tabular}[c]{@{}c@{}}Content\\ Provider\end{tabular}} \\ \hline
\textbf{Lite Apps} &  26 & 4 &   2   &    1    \\ \hline
\textbf{full Apps}  &  54   &    6   &   4  &   2      \\ \hline
\end{tabular}

\label{tab:num_component_lite_full}
\end{table}

\noindent\textbf{Results.} \textbf{Comparison at Component Level.} When comparing lite apps to their full app counterparts, it is clear that developers tend to reduce the number of interactions offered in lite apps, both in the foreground and background. This is evident in the average number of \code{Activity} (i.e., the number of UI screens) and the number of background services (i.e. \code{Service}), which are approximately half that of those found in full apps. This suggests that developers aim to offer lightweight features, reduce complexity, and improve performance in lite apps. Similar patterns can also be observed in the number of broadcast receivers and content providers in lite apps compared to full apps.

\begin{figure}[!htpb]
	\centering
	\includegraphics[width=0.5\textwidth]{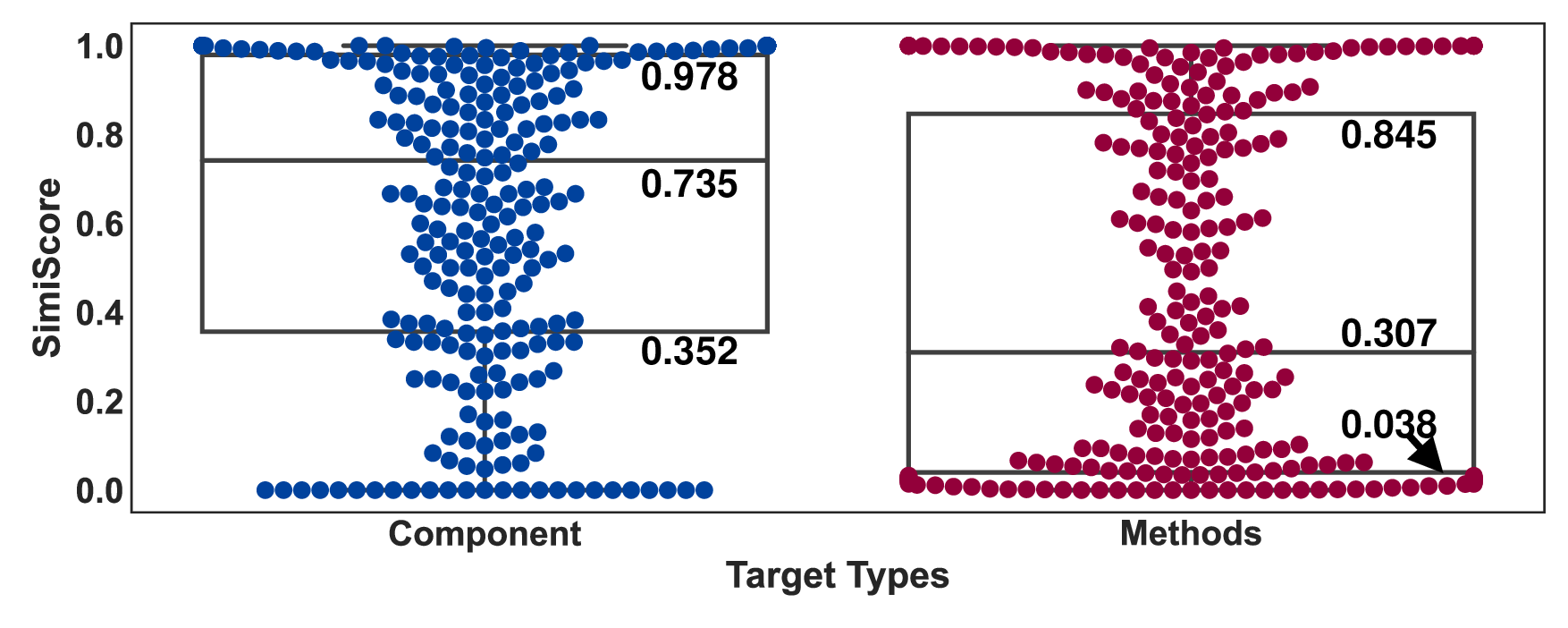}
 
	\caption{SimiScore of the Lite and full App Pairs}
 
	\label{fig:method_component_reuse}
\end{figure}

\begin{figure*}[!htpb]
	\centering
	\includegraphics[width=0.75\textwidth]{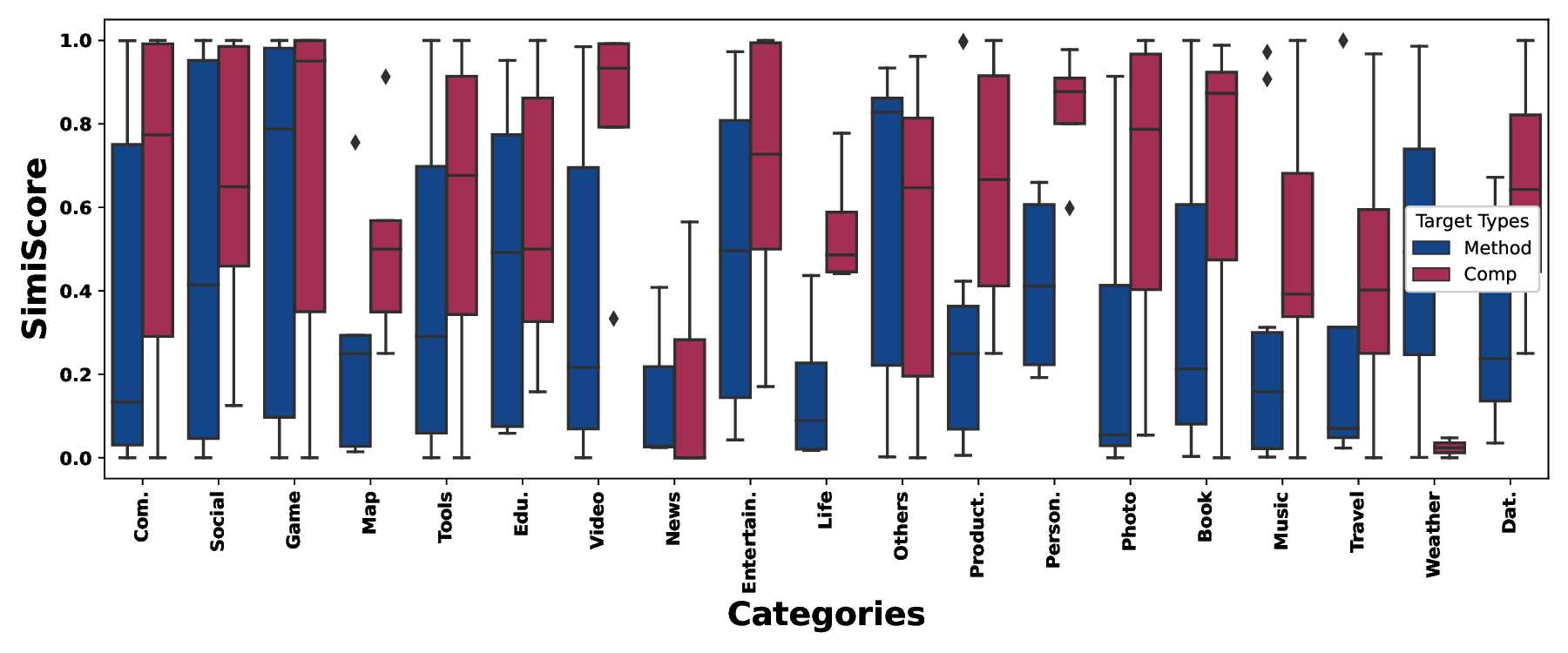}
 
	\caption{SimiScore of App Pairs by Category.}
 
	\label{fig:simscore_category}
\end{figure*}

The left side of Fig. \ref{fig:method_component_reuse} illustrates that, on average, 63.9\% of the components in lite apps are identical to those in their corresponding full apps. 50\% of lite apps have a SimiScore higher than 0.735, indicating that 73.5\% of components from the full app counterparts are reused. This highlights that a substantial number of components are transferred without modification and that lite apps often retain the same user interaction as their full app counterparts.

\noindent \textbf{Comparison at Method Level.} The right-hand side of Fig. \ref{fig:method_component_reuse} illustrates the SimiScore between lite and full app pairs in terms of methods. On average, 43\% of the methods in lite apps are identical to those in their corresponding full apps. 50\% of lite apps have a similarity score higher than 0.307, meaning that 30.7\% of methods from the full app are reused. The figure also shows that 75\% of lite apps have a similarity score higher than 0.038 (i.e., 3.8\% of methods), and 25\% of lite apps have a similarity score of 0.845 (i.e., 84.5\% of methods) are reused. 

In summary, this figure suggests that most app developers only include a few features in their lite app versions, but still maintain the majority of the components.

\begin{table}[!htpb]
\centering
\caption{Detailed Simidroid Results}
\begin{tabular}{cccc}
\hline
                    & \textbf{Identical} & \textbf{Similar}     & \textbf{New}         \\ \hline
\textbf{Components} &    52.33\%   & \multicolumn{1}{c}{1.46\%} & \multicolumn{1}{c}{46.19\%} \\ \hline
\textbf{Methods}    &      38.11\%	&	4.17\%		& 57.70\%    \\ \hline
\end{tabular}
\label{tab:detailed_simidroid}
\end{table}

Furthermore, the detailed result is shown in Tab. \ref{tab:detailed_simidroid}. It shows a detailed view of average percentage of identical, similar, new components or methods of a lite app compared with its full app counterpart. 

\noindent This suggests that (1) compared to lite apps, full apps typically have around twice as many additional features or components; and (2) a higher reuse rate of components compared to methods in lite apps suggests that developers strive to provide lightweight features in lite apps while still preserving a consistent user interface for users, in order to improve performance.

As illustrated in Fig. \ref{fig:simscore_category}, the similarity scores for components and methods vary across different app categories. However, a notable exception is the \textit{Video} and \textit{Personalization} categories, which have a consistently high component-level SimiScore compared to other categories. After thoroughly analyzing apps in the \textit{Video} and \textit{Personalization} categories, we found that most user interactions and components in both lite and full versions are identical. Specifically, developers from these two categories often do not make changes to the way users interact with the app, resulting in identical user interfaces and interactions in both versions. In the case of \textit{Video} apps, developers tend to maintain consistency in terms of user interaction, as evidenced by the identical components in both lite and full versions. In the case of \textit{Personalization} apps, the ways of interaction are relatively simple, thus there is little difference between the lite and full versions.

\begin{tcolorbox}[boxrule=1pt,boxsep=1pt,left=2pt,right=2pt,top=2pt,bottom=2pt,title=RQ2: Code Similarity]

$\bullet$ A higher reuse rate of components compared to methods in lite apps suggests that developers strive to provide lightweight features in lite apps while still preserving a consistent user interface for users, in order to improve performance;

$\bullet$ The component and method level similarities in all categories do not exhibit huge differences, except for the apps in \textit{Video} and \textit{Personalization} categories whose component-level SimiScore is significantly stable and higher than the other categories.

\end{tcolorbox}

\subsection{RQ3: App Description Similarity}
\textbf{Motivation.} In this research question, we intend to discuss whether lite apps and their corresponding full apps can have dramatically difference app descriptions.

\noindent\textbf{Methodology.} For 260 app pairs, we collect their app descriptions from Google Play Store for comparison. Next, we clean the text data by removing stop words (common words like 'the', 'is', 'in', etc.), punctuation, and converting all text to lowercase. This step helps to reduce noise in the text and makes the comparison more accurate. Furthermore, we use cosine similarity to compare app descriptions.

\begin{figure}[!htpb]
	\centering
	\includegraphics[width=0.5\textwidth]{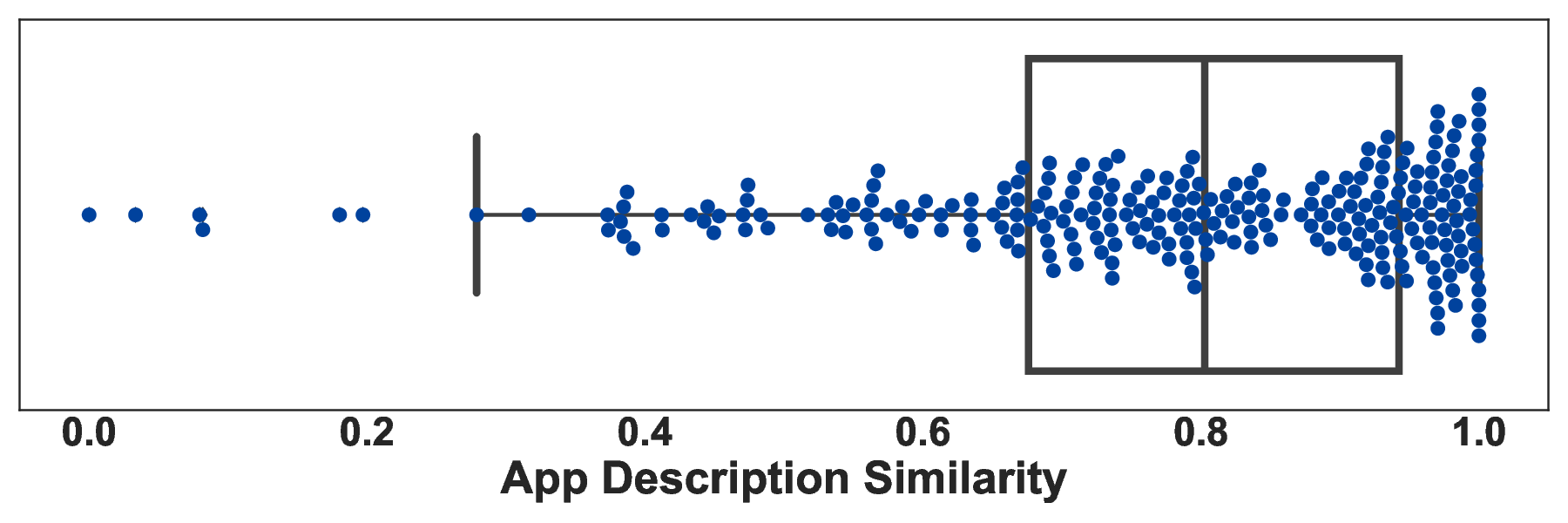}
 \vspace{-1.5em}
	\caption{Boxplot of App Description Similarity}
 \vspace{-1.5em}
	\label{fig:app_dsec_simi}
\end{figure}

\noindent\textbf{Results.} As shown in Fig. \ref{fig:app_dsec_simi}, the majority of lite apps and their corresponding apps share the similar app descriptions. On average, the lite apps and their corresponding full apps have 77.42\% similarity in terms of app descriptions. The medium value is 80.19\%, which suggests there are 50\% of apps have similarity not less than 80.19\%. There are 28 app pairs (out of 260) have similarity values less than 50\%. Next, we further investigate the lite apps and their corresponding full apps have dramatically different descriptions. As a result, we find the following clues that may lead to dramatic differences in app descriptions. 

\noindent$\bullet$ \textbf{Functionality and Features}: A lite version of an app generally has fewer features compared to its full version. Developers may cut out certain functions or remove some advanced features to simplify the user interface. Consequently, the app descriptions can be different to reflect these changes;

\noindent$\bullet$ \textbf{Target Audience}: The target audience for the lite version might be different than that of the full version. For example, a lite version may be aimed at users in emerging markets or users with older devices. Thus, the marketing language used in the description may be different to better appeal to this new audience.

\noindent$\bullet$ \textbf{Usage \& Interaction}: The way users interact with the app may also change. For example, the lite version of a social media app might be more focused on basic functions like posting status updates or photos, while the full version may allow more advanced features like live streaming. 

\begin{tcolorbox}[boxrule=1pt,boxsep=1pt,left=2pt,right=2pt,top=2pt,bottom=2pt,title=RQ3: App Description Similarity]
For the majority of app pairs, the app descriptions are similar. However, developers may design unique app descriptions for lite apps with the following possible concerns: different features/functionalities, different target audiences, and different usages/interaction modes.

\end{tcolorbox}

\subsection{RQ4: Permission, Privacy Policy Consistency}

\noindent\textbf{Motivation.} Intuitively, lite apps can be created by reworking existing full apps. However, this process can be prone to errors, which can open up opportunities for attackers. In this research question, we are interested in three aspects: (1) permission consistency: we check whether the permissions claimed are actually used in the lite app; (2) privacy policy: we examine whether (a) developers have created tailored privacy policies for lite apps, and (b) whether the privacy policies in lite apps align with the permissions they request; and (3) resources: we check whether resource files are defined but not used in the lite app.

\noindent\textbf{Methodology.} (1) For consistency of \textbf{permissions}, we leverage the PScout\cite{Au:2012} to find the mappings between APIs and permissions. We leverage PScout to find out between APIs and permissions for Android API 19. The result is reused to scan the code to determine the required permissions. Next, we compare the in-use permissions and the claimed permissions to find out over-claimed permissions; (2) For consistency of \textbf{privacy policy}, (a) for a pair of apps, we check whether the policy for the lite app is the same as the policy for the full app, to determine whether developers have created tailored privacy policies. (b) To check consistency between permissions and policies, we use AutoCog \cite{AutoCog}, which uses a learning-based algorithm to link the descriptions with permissions; and 

\begin{figure}[!htpb]
	\centering
	\includegraphics[width=0.4\textwidth]{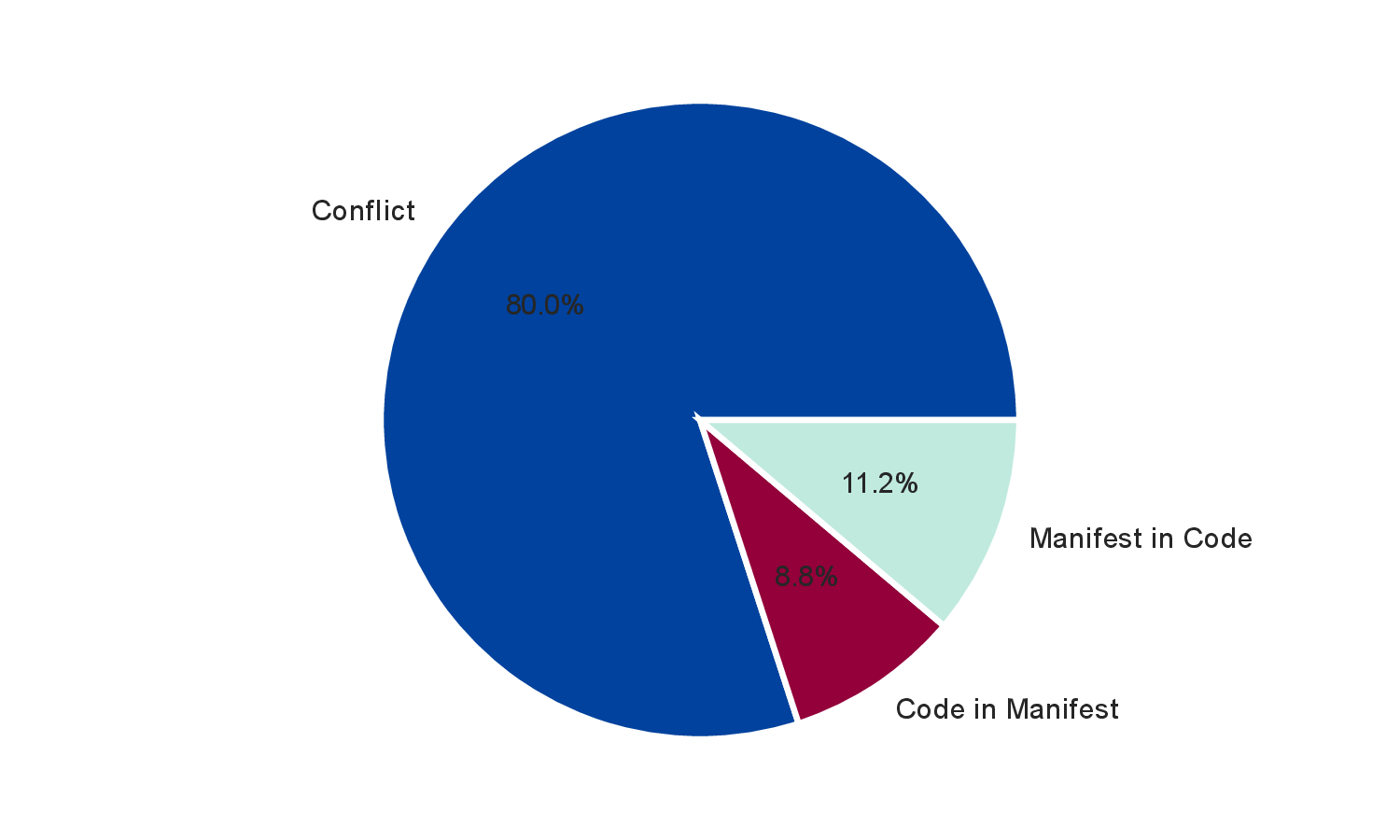}
 
	\caption{Results of Permission Consistency}
 
	\label{fig:r4permissionconsistency}
\end{figure}

\noindent\textbf{Results.}  For consistency of \textbf{permissions}, we examine all 260 lite apps and find that none of these apps maintain permission consistency. Specifically, as shown in Fig. \ref{fig:r4permissionconsistency}, there are 23 apps claim more permissions in AndroidManifest files than they used in code; 29 apps that require more permissions than they claim in the AndroidManifest files; and other 208 apps contain permission conflicts. That means, these 208 apps claim permissions that they donot use and use permissions but not claimed in the AndroidManifest files. Furthermore, we find that 89 (out of 208) apps' permission conflicts are rooted in third-party libraries; 122 (out of 208) apps' permission conflicts are rooted in main app bodies; and 3 apps' have conflicts in both main bodies and third-party libraries. 

\noindent Through out inspection of these apps, we have discovered that the possible causes for these conflicts may stem from: (1) third-party libraries: conflicts may arise if libraries used by the app request additional or different permissions than those required by the app itself, (2) codebase changes: the process of reengineering a legacy app into a lite app can result in changes to the codebase, but developers may overlook updating the corresponding permissions accordingly.

\noindent For \textbf{privacy policy} consistency, we find only 239 apps contain privacy policies. For the rest 21 apps, they do not have privacy policies published. For these 239 apps, there are only 12 apps comply with the privacy policy consistency. For the rest 227 apps, on average, each apps have 13 permissions are used but not claimed in the privacy policies, which shares a proportion of 67.5\%. It means 67.5\% of permissions are used but not but not claimed in the privacy policies. Furthermore, we find that there 163 apps out of 239 apps (68.2\%) reuse the privacy policies from their full apps. 




\begin{tcolorbox}[boxrule=1pt,boxsep=1pt,left=2pt,right=2pt,top=2pt,bottom=2pt,title=RQ4: Permission Privacy Policy and Resources Consistency]
\noindent$\bullet$ For \textbf{permission} consistency, we find that there are 23 apps claim more permissions in AndroidManifest files than they used in code; 29 apps require more permission than that they claimed in the AndroidManifest files. For \textbf{privacy policy} consistency, we find that for these 239 apps, there are only 12 apps comply with the privacy policy consistency. For the rest 227 apps, on average, each apps have 13 permissions are used but not claimed in the privacy policies, which shares a proportion of 67.5\%. 

\noindent$\bullet$ While the presence of unused resources or overclaimed permissions may not always signify a security vulnerability, it is a best practice to thoroughly review and remove them to decrease potential risks. 
\end{tcolorbox}

\subsection{RQ5: Performance Improvement with Lite App}

\noindent\textbf{Motivation.} In this RQ, we aim at evaluating whether lite apps bring performance improvement comparing to full apps from aspects: (1) size comparison; (2) startup time; (3) memory consumption; (4) CPU consumption; and (5) network requests.

\noindent\textbf{Methodology.} Specifically, to evaluate (2) \textbf{startup time}, we leverage the official the time to initial display (TTID) metric, which measures the time it takes for an application to produce its first frame, including process initialization (if a cold start), activity creation (if cold/warm), and displaying first frame. The TTID can be obtained with \code{logcat}. To evaluate (3) \textbf{memory consumption} and (4) \textbf{CPU consumption}, we leverage the adb command ``\code{adb
shell dumpsys meminfo |grep <package-name>}''
and ``\code{ps -p <pid> -o TIME}'', respectively. Note that, here ``TIME'' represents the amount of CPU time used by the process. For example, ``0:01:10'' represents the CPU time used by the process is 1 minute and 10 seconds. To evaluate (5) network requests, we perform the following steps: a) we use Soot's hierarchy analysis to find all the classes that are related to network communication with HTTP and HTTPS protocols, including, \code{java.net.URL}, \code{java.net.URLConnection}, \code{java.net.HttpURLConnection}, \newline
\code{javax.net.ssl.HttpsURLConnection}, \newline
\code{okhttp3.Request} ,and \code{okhttp3.OkHttpClient}; b) next, we iterate through each method of the class and use Soot's control flow analysis to find all the method invocations of \code{openConnection()}, \code{connect()}, \code{sendRequest()}, \code{execute()} and other similar methods that are related to making network requests. The URL can be extract with Def-Use analysis; and c) last, we extract the URLs used in the invocations, and record the URLs and the corresponding classes and methods in which they are used.


\noindent\textbf{Configuration.} To conduct our study, we use an Android emulator with the following specifications: (1) CPU/ABI: Google Play Intel Atom (x86); (2) Android API 30; and (3) 512M of memory. We utilize the Android Monkey tool to generate pseudo-random streams of user events for 15 minutes, and then calculate the memory and CPU usage.

\begin{figure}[!htpb]
	\centering
    
	\includegraphics[width=0.5\textwidth]{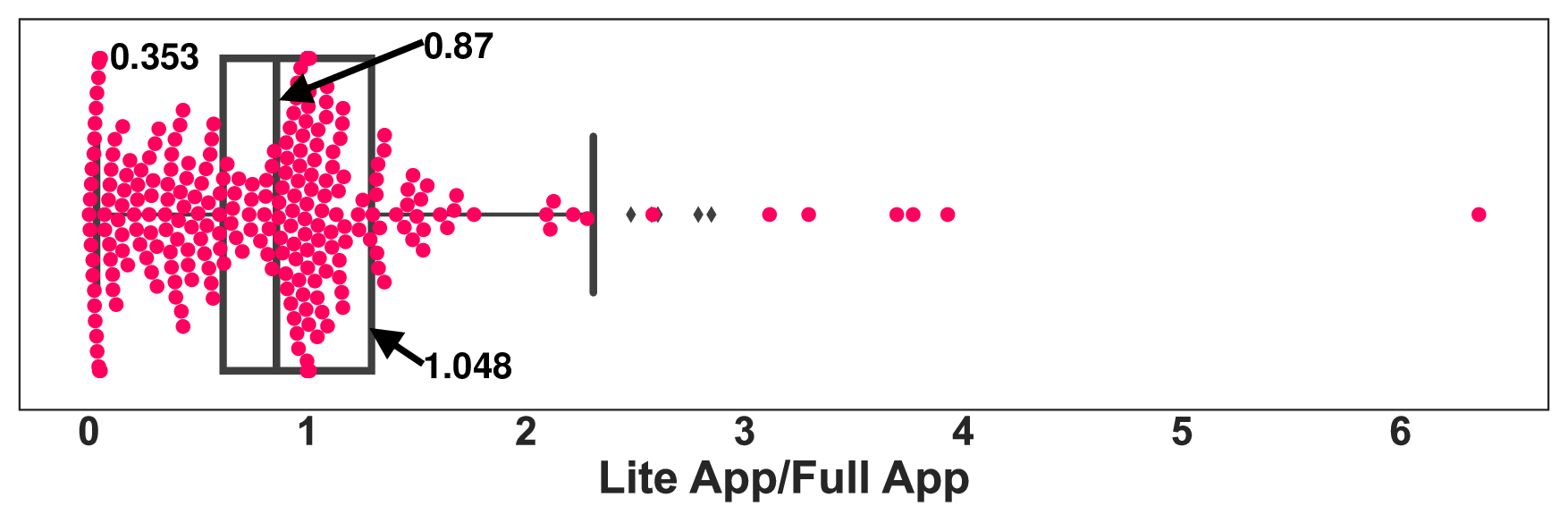}
 
	\caption{Size Comparison (lite/full) Boxplot}
 
	\label{fig:sizecomparison}
\end{figure}

\noindent\textbf{Results.} \noindent\textbf{Size Comparison.} As illustrated in Fig. \ref{fig:sizecomparison}, on average, lite apps are 83.2\% the size of their corresponding full apps. The largest proportion observed was 6.359, while the smallest was 0.001. Half of the apps have a proportion less than 0.87, with 25\% having a proportion less than 0.353 and 75\% having a proportion less than 1.048. Furthermore, 66.7\% of lite apps are smaller in size compared to their full app counterparts.

\noindent Contrary to common belief, lite apps are not always smaller in size and have fewer features compared to their full version counterparts. As shown in Fig. \ref{fig:sizecomparison}, surprisingly, around 33\% of lite apps have a larger size than their full version counterparts.

\begin{figure}[!htpb]
	\centering
	\includegraphics[width=0.5\textwidth]{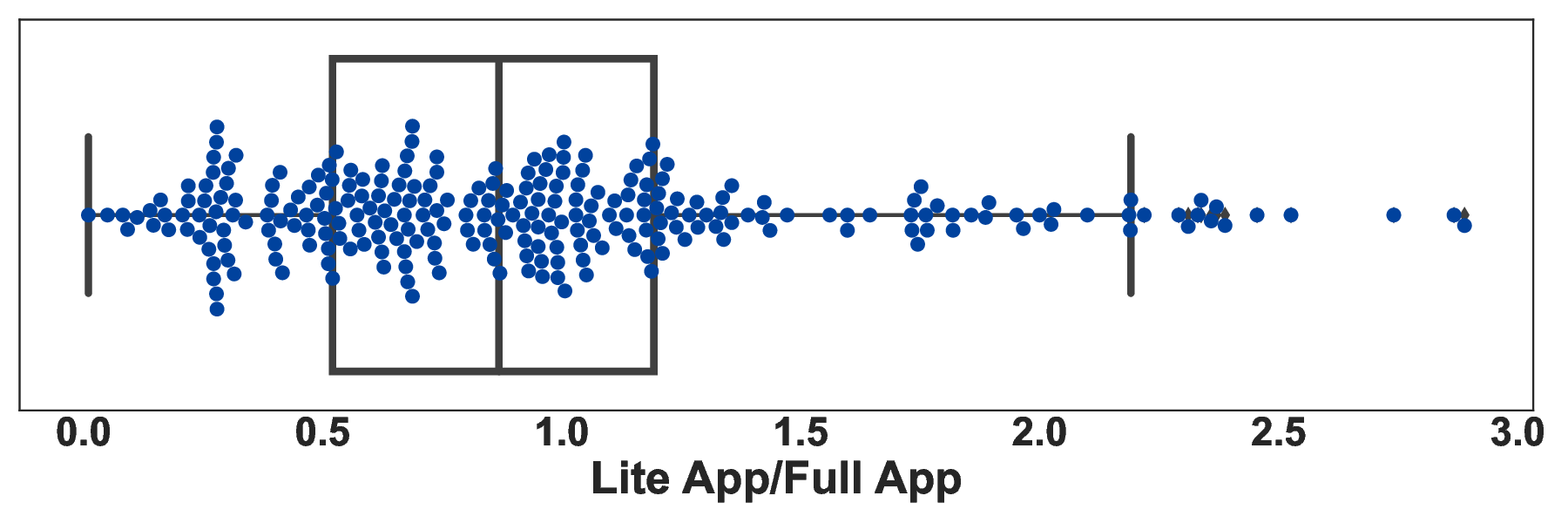}
	\caption{Startup Time Comparison (lite/full) Boxplot}
	\label{fig:startuptimecomparision}
\end{figure}

\noindent\textbf{Startup Time.} As illustrated in Fig. \ref{fig:startuptimecomparision}, on average, the lite apps have a startup time that is 94.6\% of the full app counterparts. The maximum proportion is 2.889, and the minimum proportion is 0.009. Half of the apps have a startup time that is less than 0.867 of the full app counterparts, 25\% of apps have a startup time ratio lower than 0.519, and 75\% of apps have a ratio lower than 1.191. Additionally, 61.5\% of lite apps have a startup time that is shorter than their full app counterparts.

\begin{figure}[!htpb]
	\centering
	\includegraphics[width=0.5\textwidth]{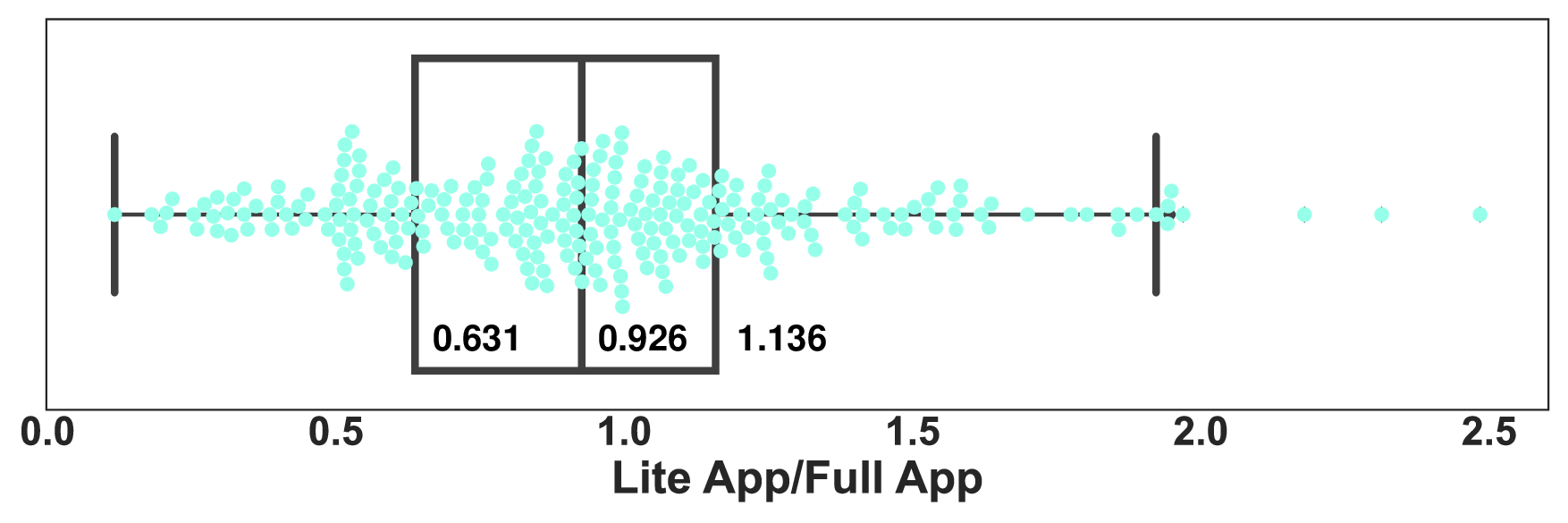}
 
	\caption{Memory Comparison (lite/full) Boxplot}
 
	\label{fig:memorycomparision}
\end{figure}

\noindent\textbf{Memory Consumption.} As seen in Fig. \ref{fig:memorycomparision}, on average, lite apps consume 0.946 times the memory of their full app counterparts. The highest proportion observed is 2.484, while the lowest is 0.116. Half of the apps have a proportion lower than 0.926, 25\% have a proportion lower than 0.631, and 75\% have a proportion lower than 1.135. This suggests that the memory consumption of 60.7\% of lite apps is not larger than that of their full app counterparts.

\begin{figure}[!htpb]
	\centering
	\includegraphics[width=0.5\textwidth]{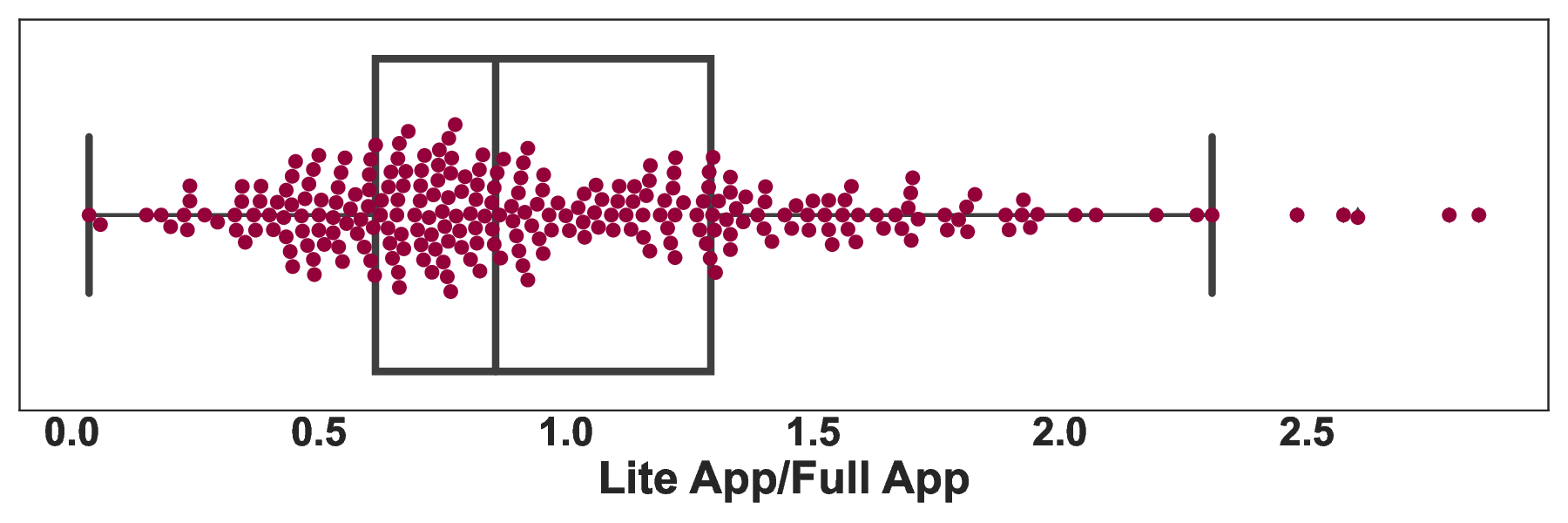}

	\caption{CPU Time Comparison (lite/full) Boxplot}

	\label{fig:cpucomparison}
\end{figure}

\noindent\textbf{CPU Consumption.} As shown in Fig. \ref{fig:cpucomparison}, on average, lite apps have a CPU consumption that is 98.5\% of their full app counterparts. The maximum proportion observed is 2.848, while the minimum is 0.03. Half of the apps have a proportion less than 0.855, 25\% of apps have a proportion less than 0.61, and 75\% of apps have a proportion less than 1.27. Additionally, 60\% of lite apps have a lower CPU consumption compared to their full app counterparts.

\begin{figure}[!htpb]
	\centering
	\includegraphics[width=0.5\textwidth]{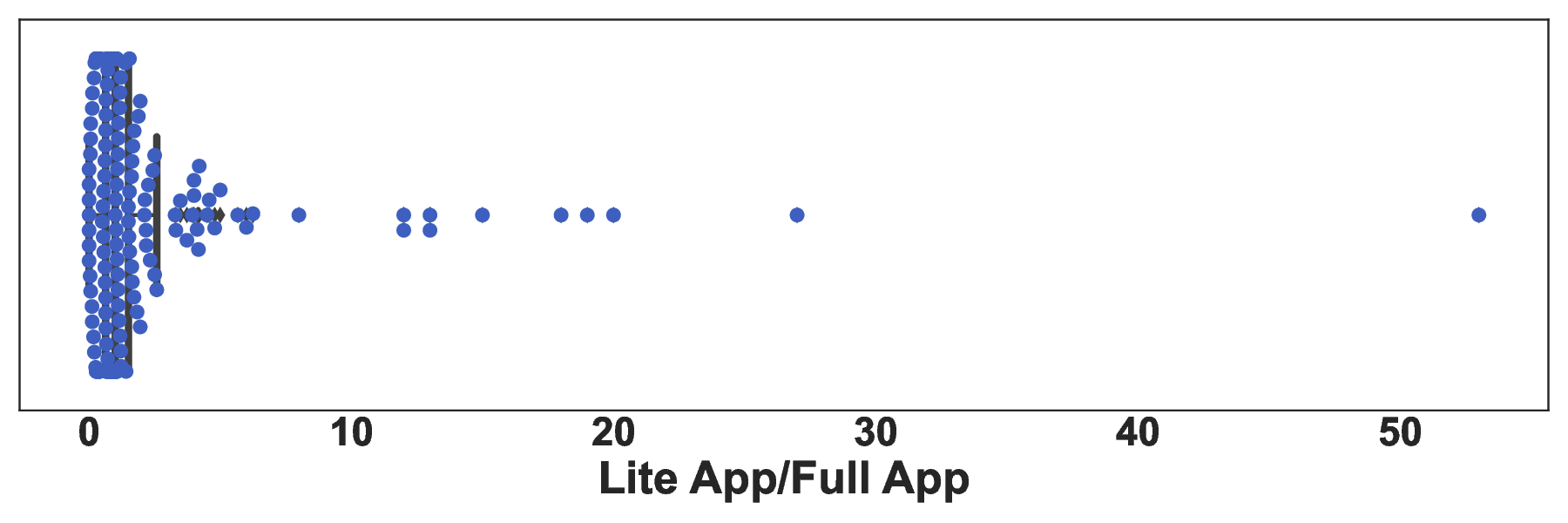}
	\caption{Network Requests Comparison(lite/full) Boxplot}
	\label{fig:networkplots}
\end{figure}

\noindent\textbf{Network Requests.} As represented in Fig. \ref{fig:networkplots}, The network requests proportion range from 0 to 53 with an average of 2.11. Furthermore, the medium value of network requests is 52 for lite apps and 53 for full apps. This suggests the lite apps send more network requests than full apps.

Furthermore, we find mere 24.6\% of lite apps exhibit improvements across all performance metrics compared to their full app counterparts. The remaining lite apps even consume more resources than their full app versions.

 
\begin{tcolorbox}[boxrule=1pt,boxsep=1pt,left=2pt,right=2pt,top=2pt,bottom=2pt,title=RQ5: Performance Improvement with Lite apps]
$\bullet$ The size proportion of lite apps to full apps ranges from 0.001 to 6.359, with an average of 0.832. The startup time proportion of lite apps to full apps ranges from 0.009 to 2.889, with an average of 0.946. The memory consumption proportion of lite apps to full apps ranges from 0.116 to 2.484, with an average of 0.946. The CPU consumption proportion of lite apps to full apps ranges from 0.03 to 2.848, with an average of 0.985; The network requests proportion of lite apps to full apps ranges from 0 to 53, with an average of 2.11.

$\bullet$ As seen in our analysis, a mere 24.6\% of lite apps exhibit improvements across all performance metrics compared to their full app counterparts. The remaining lite apps even consume more resources than their full app versions.

\end{tcolorbox}

\subsection{RQ6: Shared, Included, Excluded Features in Lite Apps}

\noindent\textbf{Motivation.} Our aim in this RQ is to understand the functional differences between lite apps and their full app counterparts. By examining which features are included and excluded in lite apps, we can gain insight into how developers decide what to include in these lightweight versions (e.g. lite apps). 

\noindent\textbf{Methodology.} We use the Diffuse toolkit \cite{diffuse} to compute the differences between the apks of lite apps and their full app counterparts. This toolkit allows us to identify the specific functionalities that are included and excluded in the lite apps. To further analyze these differences, we collect all identifiers, such as method names, variables, and class names, from the apks. We then use Latent Dirichlet Allocation (LDA) to extract relevant topics from these identifiers, with each class file being considered as a document. 

\noindent\textbf{Results.} In order to understand the feature differences between lite apps and their full app counterparts, we analyze three key areas: (1) functionality excluded in lite apps, (2) functionality included in lite apps, and (3) consistency in functionality between lite apps and full apps

\noindent\textbf{Functionality excluded in lite apps.} There are top-5 features that are used in their full app counterparts instead of lite apps:

\noindent$\bullet$ \textit{Complex data management:} In full apps, developers normally place some advanced data management features such as data syncing;

\noindent$\bullet$ \textit{Audio and video processing and support:} In full apps, full apps often include advanced audio and video processing capabilities/support with third-party libraries to process, load images and videos;

\noindent$\bullet$ \textit{Advanced support and frameworks for Games and Graphical User Interface:} Full apps often include some advanced support and frameworks (e.g., Unity Game Engine) for 3D animations, transitions, physics engines, and multiplayer support;

\noindent$\bullet$ \textit{Support for Other Devices:} Besides, we also find that full apps may offer additional support for third-party devices such as game controllers, and wearables; and

\noindent$\bullet$ \textit {Complex and Advanced Account management:} Different from lite apps, full apps often offer some advanced features for account managements, such as, multiple login options, account recovery.

\noindent\textbf{Functionality included in lite apps.} Top-5 features/functionalities that are included in lite apps, including 

\noindent$\bullet$ \textit{Lightweight and effective data processing and management:} Different from full apps, lite apps normally have more effective and lightweight data processing and management. For example, most lite apps use \code{Parcelable}, which is an interface that allows objects to be passed between activities, services, and other components in a more efficient way comparing to Serialization;

\noindent$\bullet$ \textit{Simplified concurrency and Performance Improvement:} Comparing to full apps, developers are more prone to adopt frameworks, such as Coroutines\cite{android-coroutines}, to manage asynchronous code;

\noindent$\bullet$ \textit{Custom Class/Resources Loading Mechanism} Lite apps are prone to use \code{dalvik.system.DelegateLastClassLoader} to provide a custom class loading mechanism. This gives additional benefits to lite apps, including (1) using a custom class loading mechanism can potentially improve the performance of the app by reducing the overhead of the class loading process; and (2) it allowed developers to easily add support for third-party libraries, by specifying the location of the library's class files in the classpath. The similar mechanism is used by lite apps to load resources dynamically, which can reduce the sizes of lite apps;

\noindent$\bullet$ \textit{Network connection and detection:} Lite apps often include some features to detect and check network connection status in order to determine whether it is needed to turn off background data or limit the amount of data transferred; and

\noindent$\bullet$ \textit{Reduced graphics and animations:} Lite apps often include reduced graphics and animations to reduce data usage.

\noindent\textbf{Consistency in functionality between lite apps and full apps} There are several features that are top-5 commonly found in both lite apps and their full app counterparts:

\noindent$\bullet$ \textit{Account management:} Both lite apps and full apps often include user account management features, such as, login, logout, and profile management;

\noindent$\bullet$ \textit{Notification:} The notification features in lite apps and full apps are sent to users to inform them of new content, updates, or activities within the app;

\noindent$\bullet$ \textit{Social integration:} We also find that social integration features are commonly shared by lite apps and counterpart full apps to allow users to login social account, such as Google, Twitter;

\noindent$\bullet$ \textit{Advertising:} Both lite and full apps may include advertising features to monetize their apps;

\noindent$\bullet$ \textit{App Performance Analytic:}  Both lite and full apps leverage app performance management libraries, such as, Google Analytics \cite{GoogleAnalystics}, OKHTTP3 \cite{OKHTTP}, Flurry \cite{Flurry}, to track crashes and app usages; 

\noindent In conclusion, two key insights emerge when designing features for lite apps: (1) Lite apps can function as a gateway for users to test and explore the app, potentially leading to an upgrade to the full version. (2) Performance optimization is crucial in the development of lite apps, as this will help to reach a wider audience.

\begin{tcolorbox}[boxrule=1pt,boxsep=1pt,left=2pt,right=2pt,top=2pt,bottom=2pt,title=RQ6: Shared Included and Excluded Features in Lite Apps]
\noindent$\bullet$ \textbf{Shared features in lite and full apps:} account management, notification, social integration, advertising, and app performance analytic; \textbf{Included features in lite apps:} lightweight data processing and management, simplified concurrency, customized class/resource loading, network connection and detection, and reduced graphics and animations; \textbf{Excluded features in lite apps:} complex data management, audio/video processing, advanced support for games and GUIs, cross-platform features, and advanced account management;

\end{tcolorbox}

\subsection{RQ7: Potential Security Concerns in Lite Apps}

\noindent\textbf{Motivation.} Lite apps may not have the same level of security as their full app counterparts. In this RQ, we focus on discussing the potential secure concerns from two aspects: (1) whether dangerous permissions are introduced to lite apps; (2) whether lite apps may have more sensitive data leaks than full apps. The first aspect is important as lite apps may introduce new permissions that can be exploited by attackers. The second aspect is important as lite apps may have different data handling practices than full apps, which could lead to sensitive data leaks. 

\noindent\textbf{Methodology.} Dangerous permissions in Android refer to a set of permissions that can potentially put the user's data or the device at risk. There are 38 dangerous permissions declared by Android. For (1), we check whether dangerous permissions are introduced to lite apps with AAPT. For (2), we leverage FlowDroid, which is a state-of-art taint analysis tool for Android, to detect potential privacy leaks in lite apps and full apps. Specifically, we make the following configurations in FlowDroid.

\noindent\textbf{Results.} \textbf{Dangerous Permissions:} The results of our experiments indicate that out of 260 apps, 211 (81.15\%) contain at least one dangerous permission. Furthermore, among these 211 apps, there are 83 lite apps (31.9\%) that introduce dangerous permissions that are not present in their corresponding full apps. The top introduced dangerous permissions are: GET\_ACCOUNTS (10 times), READ\_EXTERNAL\_STORAGE (9), COARSE\_LOCATION (8), READ\_PHONE\_STATE (8), WRITE\_EXTERNAL\_STORAGE (7), POST\_NOTIFICATIONS (6), FINE\_LOCATION (5), WRITE\_CALENDAR (5), and BACKGROUND\_LOCATION (4). 

It suggests a significant number of lite apps introduce dangerous permissions, which are not present in their full app counterparts, potentially putting users at risk for security and privacy breaches. Therefore, it is important for developers to be aware of the potential dangers and for users to be mindful of the permissions they grant to lite apps. Furthermore, we inspect these lite apps that introduce dangerous permissions and find the following motivations:

(1) To improve performance: developers may introduce dangerous permissions in lite apps to improve performance by accessing certain resources or data on the device. For example, developers offer complex and advanced account management support in full apps. In the lite apps, to improve performance, developers ask users to access the account through accounts Accounts Service (e.g. using a Google account), which requires the GET\_ACCOUNTS permission in lite apps for user authentication. As a result, a dangerous permission (e.g.,GET\_ACCOUNTS) is introduced in lite apps; and

(2) To collect more data from users: developers may desire to collect more data from users in order to improve their apps, promote their full apps, or target advertising more effectively.

\begin{figure}[!htpb]
	\centering
	\includegraphics[width=0.5\textwidth]{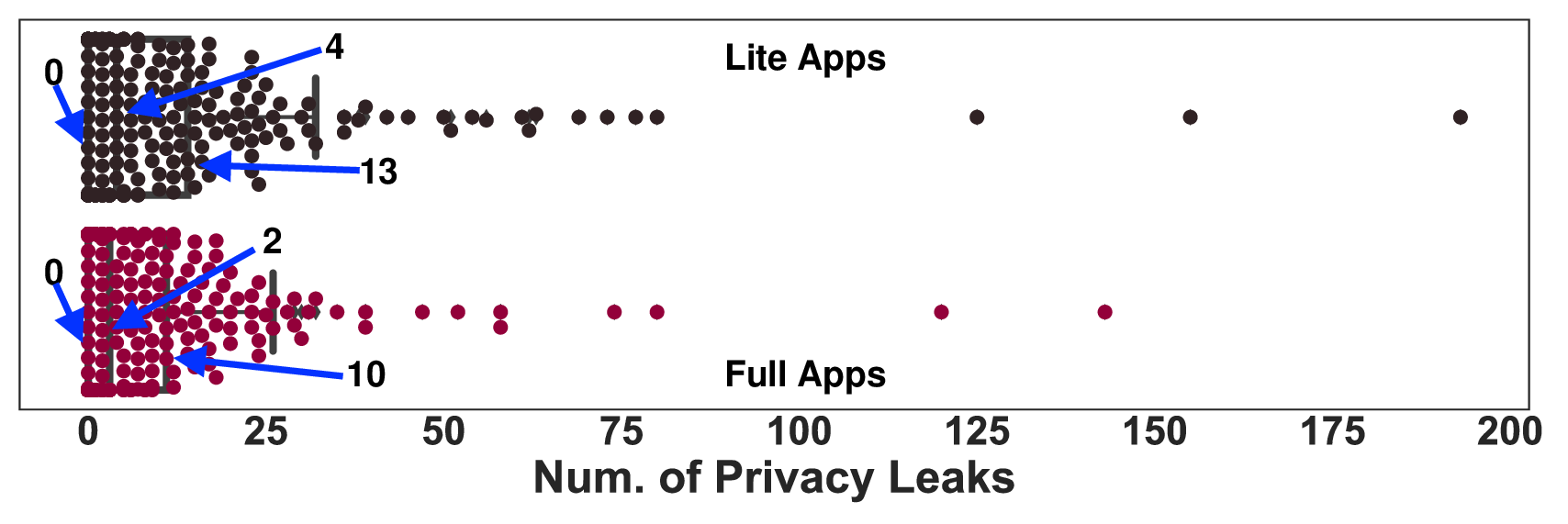}
	\caption{Number of Leaks in Apps}
	\label{fig:leaksplot}
\end{figure}

\noindent\textbf{Privacy Leaks:} Out of 520 apps (260 pairs), FlowDroid crashes on 3 apps with runtime errors. As for the rest of 257 app pairs, as represented in Fig. \ref{fig:leaksplot}, 178 lite apps out of 257 lite apps have privacy leaks, and 165 full apps contain privacy leaks. To our surprise, 161 lite apps (62.6\%) introduce new privacy leaks comparing to their corresponding full apps. On average, the lite apps have 11.4 leaks which are higher than that (8.4 leaks) in full apps. Furthermore, the medium value is 4 in lite apps, whereas that is 2 in full apps. The results indicate that lite apps offer fast and streamlined experience, particularly on devices with limited resources. However, some most lite apps have been found to potentially compromise user privacy by leaking sensitive information.

We further investigated the causes of privacy leaks in lite apps. We find the the following reasons from the samples: (1) third-party performance libraries: lite apps utilize third-party performance libraries and APIs to collect data from target users to analyze and improve their apps; (2) logging: logging libraries and APIs are used to debug, which may lead to privacy leaks that go unnoticed; and (3) implementation replacement: lite apps are designed to be lightweight, often resulting in reduced functionality, which lead to potential leaks.

\begin{tcolorbox}[boxrule=1pt,boxsep=1pt,left=2pt,right=2pt,top=2pt,bottom=2pt,title=RQ7: Potential Security Concerns in Lite Apps]

\noindent$\bullet$ There are 83 lite apps (31.9\%) that introduce dangerous permissions that are not present in their corresponding full apps. It suggests a significant number of lite apps introduce dangerous permissions, which are not present in their full app counterparts, potentially putting users at risk for security and privacy breaches.

\noindent$\bullet$ We find that 62.6\% of lite apps introduce additional privacy leaks, which are not shown in the full apps. 

\end{tcolorbox}

\section{Related Work}

\subsection{Code Clone and App Similarity} For efficiency, developers may clone the existing code into current projects. An et al.\cite{An:2017} studied the code clone from Stack Overflow to Android apps, and they find that code reuse from Stack Overflow may violate the license and cause a copyright issue. Fischer et al.\cite{Fischer:2017} found that 15.4\% of the 1.3 million Android apps they analyzed contained security-related code snippets from Stack Overflow, and 97.9\% of these code snippets contained at least one unsafe code segment. Gharehyazie et al.\cite{Gharehyazie:2019} studied the cross-project code clones in Github, they build a tool called CLONE-HUNTRESS to track code clones in Github. Due to code clones or other reasons, different apps may have similarities. Frenklach et al.\cite{Frenklach:2021} built the app similarity graph and use the app similarity graph to detect Android malware. Gonzalez et al.\cite{Gonzalez:2014} presented DroidKin, a tool to detect the similarity between Android apps. Zhou et al.\cite{Zhou:2012} build a system called DroidMOSS to detect app similarity based on fuzzy technology.

\subsection{App Analysis} There are many works on analyzing apps. Here, we only present some representative works. Arzt et al.\cite{Arzt:2014} presented the FlowDroid, a static taint analysis that has high efficiency and precision for Android apps. The Pscout\cite{Au:2012} maps Android APIs to Android permissions. A static analyzer can use the Pscout to find the permissions used by the app.  Cao et al.\cite{Cao:2015} built the EdgeMiner to detect implicit control flow transitions in the Android framework automatically. Li et al.\cite{Li:2015} presented the Violist, a string analysis framework to extract the string value from the target APIs in Android apps. The GUI analysis for Android apps mainly contributes to the understanding of apps' testing and behaviors \cite{Wang:2018,Guo:2020,Mahmood:2014,Stoat,A3E,Sapienz,yang:icse15,yang:ase2018,Rountev:2014}. Specifically, Rountev and Yang \cite{Rountev:2014} proposed a static object reference analysis approach for GUI Objects in apps. The static object reference analysis approach employs a constraint graph to model the flow of GUI objects and the hierarchical structure of these objects. Amalfitano et al. \cite{Amalfitano:2012} proposed the AndroidRipper, a UI driven ripper that tests Android apps via their GUI. Wang et al \cite{Wang:2018} presented a novel analysis named \textit{input context analysis} on an app's GUI to detect whether there exist privacy leaks for user-entered data for a given app and determining whether such leakage can lead to a violation of the privacy policy claimed. Yang et al. \cite{yang:icse15,yang:ase2018} proposed a static analysis approach to build the window transaction graph for Android apps. Guo et al \cite{Guo:2020} first conduct a systematical study on learning the limitations of existing app GUI fuzzers. Then, they proposed a novel GUI fuzzing approach by leveraging additional dependencies relations between UI objects. There are also some works explore automatically testing Android apps by exploring pages in the app under test \cite{Mahmood:2014,Stoat,A3E,Sapienz}.

\subsection{App Debloating} The bloated apps consume more resources and may have more security vulnerabilities compared to full apps. McDaniel et al.\cite{McDaniel:2012} explored the bloatware in smartphones. Elahi et al.\cite{Elahi:2020} found an average of 172 bloated apps in the smartphone inspected; most of them can access sensitive data. To detect and remove software bloat in apps, Tang et al.\cite{Tang:2021} proposed the XDebloat which can help developers remove unwanted features and transform a bloated app into an instant app or an app bundle. Jiang et al.\cite{Jiang:18} studied bloated Android apps, and they built a prototype called RedDroid to identify the bloated app and remove dead code in apps.

\section{Conclusion}
In this study, we compare and analyze the differences between full apps and their corresponding lite versions. Our analysis includes a thorough examination of resource similarity, code similarity, permission consistency, privacy policy consistency, resource consistency, performance comparison, and feature practices in lite apps. The results of our study indicate that many existing lite apps do not achieve their intended goals, such as being smaller in size, faster performance, and consuming less data. Additionally, our study highlights the potential security risks that lite apps may pose. To assist app developers in building effective lite apps and debloating tools, we compare and discuss the functionalities in both lite and full apps. 

\section{Data Availability}
The experimental results, raw data, and apps are available at: \url{https://sites.google.com/view/liteapps/home}. 


\balance

\bibliographystyle{IEEEtran}
\bibliography{ref}

\begin{thebibliography}{10}
\providecommand{\url}[1]{#1}
\csname url@samestyle\endcsname
\providecommand{\newblock}{\relax}
\providecommand{\bibinfo}[2]{#2}
\providecommand{\BIBentrySTDinterwordspacing}{\spaceskip=0pt\relax}
\providecommand{\BIBentryALTinterwordstretchfactor}{4}
\providecommand{\BIBentryALTinterwordspacing}{\spaceskip=\fontdimen2\font plus
\BIBentryALTinterwordstretchfactor\fontdimen3\font minus \fontdimen4\font\relax}
\providecommand{\BIBforeignlanguage}[2]{{%
\expandafter\ifx\csname l@#1\endcsname\relax
\typeout{** WARNING: IEEEtran.bst: No hyphenation pattern has been}%
\typeout{** loaded for the language `#1'. Using the pattern for}%
\typeout{** the default language instead.}%
\else
\language=\csname l@#1\endcsname
\fi
#2}}
\providecommand{\BIBdecl}{\relax}
\BIBdecl

\bibitem{appbrain}
Appbrain, ``Appbrain,'' 2022, https://www.appbrain.com/.

\bibitem{Huang:2017}
J.~Huang, Y.~Aafer, D.~Perry, X.~Zhang, and C.~Tian, ``Ui driven android application reduction,'' in \emph{2017 32nd IEEE/ACM International Conference on Automated Software Engineering (ASE)}, 2017, pp. 286--296.

\bibitem{Tang:2021}
Y.~Tang, H.~Zhou, X.~Luo, T.~Chen, H.~Wang, Z.~Xu, and Y.~Cai, ``Xdebloat: Towards automated feature-oriented app debloating,'' \emph{IEEE Transactions on Software Engineering}, pp. 1--19, 2021.

\bibitem{Bhattacharya:13}
S.~Bhattacharya, K.~Gopinath, and M.~G. Nanda, ``Combining concern input with program analysis for bloat detection,'' in \emph{Proceedings of OOPSLA}.\hskip 1em plus 0.5em minus 0.4em\relax Association for Computing Machinery, 2013, p. 745–764.

\bibitem{Jiang:18}
Y.~Jiang, Q.~Bao, S.~Wang, X.~Liu, and D.~Wu, ``Reddroid: Android application redundancy customization based on static analysis,'' in \emph{Proceedings of ISSRE}, 2018, pp. 189--199.

\bibitem{appbundle}
Google, ``App bundle,'' \url{https://developer.android.com/platform/technology/app-bundle/}, 2022.

\bibitem{googleplay}
------, ``Google play,'' https://play.google.com/store/apps, 2022.

\bibitem{AAPT}
Android, ``Aapt,'' https://developer.android.com/studio/command-line/aapt2, 2022.

\bibitem{Li:2017}
L.~Li, T.~F. Bissyand{\'e}, and J.~Klein, ``Simidroid: Identifying and explaining similarities in android apps,'' in \emph{2017 IEEE Trustcom/BigDataSE/ICESS}, 2017, pp. 136--143.

\bibitem{Soot}
Soot, ``Soot - a java optimization framework,'' \url{https://github.com/soot-oss/soot}, 2022.

\bibitem{Au:2012}
K.~W.~Y. Au, Y.~F. Zhou, Z.~Huang, and D.~Lie, ``Pscout: analyzing the android permission specification,'' in \emph{Proceedings of the 2012 ACM conference on Computer and communications security}, 2012, pp. 217--228.

\bibitem{AutoCog}
Z.~Qu, V.~Rastogi, X.~Zhang, Y.~Chen, T.~Zhu, and Z.~Chen, ``Autocog: Measuring the description-to-permission fidelity in android applications,'' in \emph{Proceedings of CCS}, 2014, pp. 1354--1365.

\bibitem{diffuse}
Diffuse, ``Diffuse is a tool for diffing apks, aabs, aars, and jars,'' 2022, https://github.com/JakeWharton/diffuse.

\bibitem{android-coroutines}
Android, ``Coroutines on android,'' 2022, https://developer.android.com/kotlin/coroutines.

\bibitem{GoogleAnalystics}
GoogleAnalystics, ``Googleanalystics,'' 2022, https://analytics.google.com/analytics/web/.

\bibitem{OKHTTP}
OKHTTP3, ``Okhttp,'' 2022, https://square.github.io/okhttp/.

\bibitem{Flurry}
Flurry, ``Flurry: Mobile app analytics platform for android and ios,'' 2022, https://www.flurry.com/.

\bibitem{An:2017}
L.~An, O.~Mlouki, F.~Khomh, and G.~Antoniol, ``Stack overflow: A code laundering platform?'' in \emph{2017 IEEE 24th International Conference on Software Analysis, Evolution and Reengineering (SANER)}.\hskip 1em plus 0.5em minus 0.4em\relax IEEE, 2017, pp. 283--293.

\bibitem{Fischer:2017}
F.~Fischer, K.~B{\"o}ttinger, H.~Xiao, C.~Stransky, Y.~Acar, M.~Backes, and S.~Fahl, ``Stack overflow considered harmful? the impact of copy\&paste on android application security,'' in \emph{2017 IEEE Symposium on Security and Privacy (SP)}.\hskip 1em plus 0.5em minus 0.4em\relax IEEE, 2017, pp. 121--136.

\bibitem{Gharehyazie:2019}
M.~Gharehyazie, B.~Ray, M.~Keshani, M.~S. Zavosht, A.~Heydarnoori, and V.~Filkov, ``Cross-project code clones in github,'' \emph{Empirical Software Engineering}, vol.~24, no.~3, pp. 1538--1573, 2019.

\bibitem{Frenklach:2021}
T.~Frenklach, D.~Cohen, A.~Shabtai, and R.~Puzis, ``Android malware detection via an app similarity graph,'' \emph{Computers \& Security}, vol. 109, p. 102386, 2021.

\bibitem{Gonzalez:2014}
H.~Gonzalez, N.~Stakhanova, and A.~A. Ghorbani, ``Droidkin: Lightweight detection of android apps similarity,'' in \emph{International Conference on Security and Privacy in Communication Networks}.\hskip 1em plus 0.5em minus 0.4em\relax Springer, 2014, pp. 436--453.

\bibitem{Zhou:2012}
W.~Zhou, Y.~Zhou, X.~Jiang, and P.~Ning, ``Detecting repackaged smartphone applications in third-party android marketplaces,'' in \emph{Proceedings of the second ACM conference on Data and Application Security and Privacy}, 2012, pp. 317--326.

\bibitem{Arzt:2014}
S.~Arzt, S.~Rasthofer, C.~Fritz, E.~Bodden, A.~Bartel, J.~Klein, Y.~Le~Traon, D.~Octeau, and P.~McDaniel, ``Flowdroid: Precise context, flow, field, object-sensitive and lifecycle-aware taint analysis for android apps,'' in \emph{Proceedings of PLDI}, 2014, p. 259–269.

\bibitem{Cao:2015}
Y.~Cao, Y.~Fratantonio, A.~Bianchi, M.~Egele, C.~Kruegel, G.~Vigna, and Y.~Chen, ``Edgeminer: Automatically detecting implicit control flow transitions through the android framework.'' in \emph{NDSS}, 2015.

\bibitem{Li:2015}
D.~Li, Y.~Lyu, M.~Wan, and W.~G. Halfond, ``String analysis for java and android applications,'' in \emph{Proceedings of the 2015 10th Joint Meeting on Foundations of Software Engineering}, 2015, pp. 661--672.

\bibitem{Wang:2018}
X.~Wang, X.~Qin, M.~B. Hosseini, R.~Slavin, T.~D. Breaux, and J.~Niu, ``Guileak: Tracing privacy policy claims on user input data for android applications,'' in \emph{Proceedings of ICSE}, 2018, p. 37–47.

\bibitem{Guo:2020}
W.~Guo, L.~Shen, T.~Su, X.~Peng, and W.~Xie, ``Improving automated gui exploration of android apps via static dependency analysis,'' in \emph{Proceedings of ICSME}, 2020, pp. 557--568.

\bibitem{Mahmood:2014}
R.~Mahmood, N.~Mirzaei, and S.~Malek, ``Evodroid: Segmented evolutionary testing of android apps,'' in \emph{Proceedings of ESEC/FSE}, 2014, p. 599–609.

\bibitem{Stoat}
T.~Su, G.~Meng, Y.~Chen, K.~Wu, W.~Yang, Y.~Yao, G.~Pu, Y.~Liu, and Z.~Su, ``Guided, stochastic model-based gui testing of android apps,'' in \emph{Proceedings of ESEC/FSE}, 2017, p. 245–256.

\bibitem{A3E}
T.~Azim and I.~Neamtiu, ``Targeted and depth-first exploration for systematic testing of android apps,'' \emph{SIGPLAN Not.}, vol.~48, no.~10, p. 641–660, 2013.

\bibitem{Sapienz}
K.~Mao, M.~Harman, and Y.~Jia, ``Sapienz: Multi-objective automated testing for android applications,'' in \emph{Proceedings of ISSTA}, 2016, p. 94–105.

\bibitem{yang:icse15}
S.~Yang, D.~Yan, H.~Wu, Y.~Wang, and A.~Rountev, ``Static control-flow analysis of user-driven callbacks in {A}ndroid applications,'' in \emph{Proceedings of ICSE}, 2015, pp. 89--99.

\bibitem{yang:ase2018}
S.~Yang, H.~Wu, H.~Zhang, Y.~Wang, C.~Swaminathan, D.~Yan, and A.~Rountev, ``Static window transition graphs for android,'' \emph{Automated Software Engineering}, vol.~25, no.~4, pp. 833--873, 2018.

\bibitem{Rountev:2014}
A.~Rountev and D.~Yan, ``Static reference analysis for gui objects in android software,'' in \emph{Proceedings of CGO}, 2014, p. 143–153.

\bibitem{Amalfitano:2012}
D.~Amalfitano, A.~R. Fasolino, P.~Tramontana, S.~De~Carmine, and A.~M. Memon, ``Using gui ripping for automated testing of android applications,'' in \emph{Proceedings of ASE}, 2012, pp. 258--261.

\bibitem{McDaniel:2012}
P.~McDaniel, ``Bloatware comes to the smartphone,'' \emph{IEEE Security \& Privacy}, vol.~10, no.~4, pp. 85--87, 2012.

\bibitem{Elahi:2020}
H.~Elahi, G.~Wang, and J.~Chen, ``Pleasure or pain? an evaluation of the costs and utilities of bloatware applications in android smartphones,'' \emph{Journal of Network and Computer Applications}, vol. 157, p. 102578, 2020.

\end{thebibliography}

\end{document}